\documentclass[10pt,twocolumn,prd,superscriptaddress,showpacs,floatfix,preprintnumbers,nofootinbib]{revtex4}   
\usepackage{epsfig}   
\usepackage{multirow}

\long\def\ignore#1{}

\usepackage[dvips]{color}   
\definecolor{Black}{named}{Black}   
\definecolor{Blue}{named}{Blue}   
\newcommand{\msqs}{\Delta m^2_{\odot}}   
\newcommand{\msqstilde}{\Delta{\tilde{m}}^2_{\odot}}   
\def\beq{\begin{equation}}   
\def\eeq{\end{equation}}

\begin{document}   
   
\title{Identifying neutrino mass hierarchy at extremely small   
$\theta_{13}$ \\   
 through Earth matter effects in a supernova signal}   
   
\author{Basudeb Dasgupta}   
\affiliation{Tata Institute of Fundamental Research, Homi Bhabha   
Road, Mumbai 400005, India}   
   
\author{Amol Dighe}   
\affiliation{Tata Institute of Fundamental Research, Homi Bhabha   
Road, Mumbai 400005, India}   
   
\author{Alessandro Mirizzi}   
\affiliation{Max-Planck-Institut f\"ur Physik   
(Werner-Heisenberg-Institut), F\"ohringer Ring 6, 80805 M\"unchen,   
Germany}   
   
\preprint{MPP-2008-14,TIFR/TH/08-04}   
   
%%%%%%%%%%%%%%%%%%%%%%%%%%%%%%%%%%%%%%%%%%%%%%%%%%%%%%%%%%%%%%%%%%%%%%   
\begin{abstract}   
 
Collective neutrino flavor transformations deep inside a supernova 
are sensitive to the neutrino mass hierarchy even at 
extremely small values of $\theta_{13}$. 
Exploiting this effect, we show that comparison of the  
antineutrino signals from a galactic supernova in 
two megaton class water Cherenkov detectors, 
one of which is shadowed by the Earth, 
will enable us to distinguish between the hierarchies if  
$\sin^2 \theta_{13} \lesssim 10^{-5}$,  
where long baseline neutrino experiments would be ineffectual.

\end{abstract}   
%%%%%%%%%%%%%%%%%%%%%%%%%%%%%%%%%%%%%%%%%%%%%%%%%%%%%%%%%%%%%%%%%%%%%%   
   
\pacs{14.60.Pq, 97.60.Bw}

\maketitle   
   
%%%%%%%%%%%%%%%%%%%%%%%%%%%%%%%%%%%%%%%%%%%%%%%%%%%%%%%%%%%%%%%%%%%%%%   
 
After many years of experiments on atmospheric,  
solar and terrestrial neutrinos, our knowledge of neutrino masses and mixing    
has grown by leaps and   
bounds~\cite{Bandyopadhyay:2008va,GonzalezGarcia:2007ib,Fogli:2006yq,Maltoni:2004ei}.   
The data is now described satisfactorily in the 3-neutrino oscillation  
framework defined by two mass squared differences $\msqs$ and   
 $\Delta m^2_{\rm atm}$, three mixing angles $\theta_{12}$,    
$\theta_{23}$ and $\theta_{13}$, and the CP-violating phase $\delta$.   
The parameters $\theta_{13}$ and $|\Delta m^2_{\rm atm}|$  
are determined by atmospheric neutrino experiments and 
long baseline experiments, while $\msqs$ and  
$\theta_{12}$ are determined by solar and reactor experiments.   
Current data on neutrino oscillations do not determine   
the sign of $\Delta m^2_{\rm atm}$.  
One refers to $\Delta m^2_{\rm atm}>0$ as normal mass hierarchy and  
$\Delta m^2_{\rm atm}<0$ as inverted mass hierarchy.   
For $\theta_{13}$ there is only an upper bound   
$\sin^2  \theta_{13} < 0.02$ at 90 \% C.L. from reactor experiments.  
The phase $\delta$ is completely unknown.

The primary goals of next generation neutrino experiments include   
the measurement of the leptonic mixing angle $\theta_{13}$,  
and the determination of  the neutrino mass hierarchy.   
The present and near-future long baseline and reactor experiments are expected    
to probe $\theta_{13}$ down to $\sin^2\theta_{13}\lesssim 10^{-2}-10^{-3}$   
(see, e.g.,~\cite{Bandyopadhyay:2007kx}).   
Over the next decade, if a neutrino factory or a superbeam facility is built,   
the reach could get extended to $\sin^2 \theta_{13} \lesssim 10^{-4}-  
10^{-5}$~\cite{Bandyopadhyay:2007kx,nufact}.    
Several studies have been performed to ascertain the optimal strategy   
for determining the neutrino mass hierarchy if $\theta_{13}$ is  
large enough to be measured at the upcoming experiments    
(see, e.g.,~\cite{Gandhi:2007td,Schwetz:2007py,Learned:2006wy,Minakata:2006gq}). 
However, if $\theta_{13}$ is small, then the determination of the  
mass hierarchy becomes very challenging even with  
next-to-next generation neutrino factory  
experiments~\cite{deGouvea:2005mi,Bross:2007ts}.

In this Letter, we propose a new astrophysical method for determining   
the neutrino mass   
hierarchy, which works for extremely small values of $\theta_{13}$  
where the standard oscillation experiments fail.    
This method makes use of the Earth matter effects on the neutrino signal   
from a galactic supernova (SN). 
   
Mikheyev-Smirnov-Wolfenstein (MSW) flavor  
conversions~\cite{Wolfenstein:1977ue,Mikheev:1986gs}    
in a SN envelope are known to mix the primary neutrino fluxes.
This mixing is crucially dependent on $\theta_{13}$ and the neutrino  
mass hierarchy.    
In addition to flavor conversions in the star, neutrinos arriving  
from a SN can undergo    
oscillations inside the Earth before being observed at a shadowed    
detector~\cite{Dighe:1999bi,Lunardini:2001pb,Takahashi:2000it}.    
These Earth effects associated with supernova neutrinos have been  
studied extensively as a powerful tool to probe neutrino mass hierarchy    
for $\sin^2 \theta_{13}  
\gtrsim10^{-3}$~\cite{Dighe:2003be,Dighe:2003jg,Lunardini:2003eh,Dighe:2003vm}. 
For $\sin^2 \theta_{13} \lesssim 10^{-5}$,  
the traditional analysis of neutrino flavor conversions in SN,  
where only MSW transitions were taken into account, predicts no    
hierarchy dependence of SN neutrino spectra arriving at  
the Earth~\cite{Dighe:2007ks}.    
Therefore, observation of the Earth matter effect cannot determine  
the neutrino  mass hierarchy in that case.  
This picture is profoundly modified  
when one considers the new emerging paradigm of  
collective effects on supernova neutrino flavor  
conversions.  
    
Recently, it was pointed  
out~\cite{Duan:2006an,Hannestad:2006nj,Fogli:2007bk} that    
large neutrino density near the neutrinosphere results in significant coherent    
neutrino--neutrino forward scattering,   
which gives rise to collective neutrino flavor oscillations inside the SN.   
Three-flavor analysis of collective effects has now been carried    
out~\cite{Dasgupta:2007ws}, which allows us to    
characterize collective oscillation effects and to write down the    
flavor spectra of neutrinos and antineutrinos arriving at the Earth.   
Following~\cite{Dasgupta:2007ws}, we work in the modified flavor  
basis $(\nu_e,\nu_x,\nu_y)$, defined such that   
$(\nu_e,\nu_x,\nu_y)= R_{23}^{\dagger}(\theta_{23})(\nu_e,\nu_\mu,\nu_\tau)$,  
where $R_{23}$ is the 2-3 rotation matrix.  
As a result of these collective oscillations, there are collective pair conversions    
$\nu_e {\bar \nu_e}  \leftrightarrow \nu_y {\bar \nu_y}$    
within the first ${\cal O}(100)$~km~\cite{Hannestad:2006nj}.

The manifestation of these flavor transformations depends on the primary spectra of neutrinos.
 In typical supernova models one finds a hierarchy of number fluxes $N_{\nu_e} >
N_{{\bar \nu}_e}> N_{\nu_x}=N_{{\bar \nu}_x}$~\cite{Keil:2002in,Totani:1997vj}. 
Even if it is not obvious that this hierarchy is maintained also at late times,
 in the following we will assume it as our benchmark.
 This scenario has been extensively studied analytically as well as
numerically, and gives straightforward predictions for neutrino flavor
conversions.
In the normal hierarchy the spectra remain unaffected by collective oscillations. In the inverted hierarchy, for any non-zero value   
of $\theta_{13}$ such that the adiabatic solution in \cite{Raffelt:2007cb}
is valid, the end of collective oscillations   
is marked by a complete exchange of the $e$ and $y$ flavor spectra  
for ${\bar\nu}$.
The $\nu$ spectra also get swapped, however only above
a characteristic energy,
giving rise to a split in the spectrum~\cite{Raffelt:2007cb,Duan:2007fw}.
This solution is valid for extremely small $\theta_{13}$, as long
 as bipolar oscillations develop sufficiently~\cite{Duan:2007bt} and the evolution is adiabatic.
 We have checked that the adiabatic solution remains valid at values of $\theta_{13}$
 that are as low as $10^{-10}$, for typical SN neutrino density profiles. 
As a consequence, neutrino fluxes which are further processed by MSW matter
 effects are  significantly different for the two hierarchies, even for extremely small
 $\theta_{13}$ values. 
This sensitivity presents a novel possibility to determine the mass hierarchy at small $\theta_{13}$.
 We must remark that qualitatively different primary neutrino spectra and/or yet undiscovered 
flavor effects may yield different predictions 
for flavor conversion and the analysis will have to be repeated appropriately.

Here we concentrate on the $\bar{\nu}_e$ spectra observable    
through inverse beta decay reactions ${\bar \nu}_e +p\to n+e^+$
at water Cherenkov detectors.     
In inverted hierarchy, MSW matter effects  in SN envelope  
are characterized in terms of the level-crossing  
probability $P_H$~\cite{Dighe:1999bi,Fogli:2001pm} of antineutrinos,  
which is in general a function  of the  neutrino energy and $\theta_{13}$.   
In the following, we consider  two extreme limits,   
$P_H\simeq0$ when $\sin^2 \theta_{13} \gtrsim 10^{-3}$ (``large''),  
and $P_H \simeq 1$ when $\sin^2 \theta_{13}\lesssim 10^{-5}$ 
(``small'').

While propagating through the Earth, the 
$\bar{\nu}_e$ and $\bar{\nu}_x$ spectra partially mix.
The neutrino fluxes $F_{\nu}$ at the Earth surface for
normal hierarchy, as well as for inverted hierarchy with  
large $\theta_{13}$, are given in terms of the the primary fluxes $F^0_{\nu}$ by
\begin{eqnarray}  
F_{\bar{e}} & = & \cos^2 \theta_{12} F^0_{\bar{e}} +  
\sin^2 \theta_{12} F^0_{\bar{x}} \; , \nonumber \\
F_{\bar{x}} & = & \sin^2 \theta_{12} F^0_{\bar{e}} +  
\cos^2 \theta_{12} F^0_{\bar{x}} \; . 
\label{nh,ihL}  
\end{eqnarray}

For inverted hierarchy with small $\theta_{13}$, we have  
\begin{eqnarray}  
F_{\bar{e}}= \cos^2 \theta_{12} F^0_{\bar{y}} +  
\sin^2 \theta_{12} F^0_{\bar{x}}  \approx F^0_{\bar{x}} \;, \nonumber \\
F_{\bar{x}}= \sin^2 \theta_{12} F^0_{\bar{y}} +  
\cos^2 \theta_{12} F^0_{\bar{x}}  \approx F^0_{\bar{x}} \;.
\label{ihS}  
\end{eqnarray}  
    
Earth effect can be taken into account by just mapping $\cos^2 \theta_{12}   
\to P({\bar \nu}_1 \to {\bar \nu_e})$ and $\sin^2 \theta_{12} \to   
1-  P({\bar \nu}_1 \to {\bar \nu_e})$, where $P({\bar \nu}_1 \to {\bar \nu_e})$   
is the probability that a state entering the Earth as  
mass eigenstate ${\bar\nu}_1$ is detected   
as ${\bar\nu}_e$ at the detector.    
   
From Eqs.~(\ref{nh,ihL}) and (\ref{ihS}), one expects to observe Earth   
matter effect in normal hierarchy independently of $\theta_{13}$,   
while in inverted hierarchy it is  expected only at large $\theta_{13}$.   
For small $\theta_{13}$ and inverted hierarchy,   
the ${\bar\nu}_e$ spectrum arriving at the Earth is  
identical to the $\bar{\nu}_x$ spectrum arriving at the Earth,  
so any oscillation effect among them is unobservable.   
This implies that if next generation neutrino experiments   
bound $\theta_{13}$ to be small, from    
the (non)observation of Earth matter effect we could identify 
the neutrino mass hierarchy.    
   
A strategy to observe Earth matter signatures in neutrino oscillations is to 
 compare the signal at two detectors.   
The difference between the ${\bar \nu}_e$ flux $F_{\bar e}^D$ at a shadowed    
detector and the ${\bar \nu}_e$ flux $F_{\bar e}$ at a detector that is not    
shadowed by the Earth   
can be written as   
%..................................................   
\begin{equation}   
\Delta F = F_{\bar e}^D-F_{\bar e}=  f_{\rm {reg}}(F_{\bar e}^0-   
F^0_{\bar x}) \,\ ,   
\label{eq:deltaf}   
\end{equation}   
%.................................................   
for normal hierarchy as well as for inverted hierarchy with   
large $\theta_{13}$.  
Here $f_{\rm reg}= P({\bar \nu}_1 \to {\bar \nu_e}) - \cos^2 \theta_{12}$    
is the Earth regeneration factor.   
In inverted hierarchy for    
small $\theta_{13}$, we get $\Delta F=0$.    
If the $\bar{\nu}$ trajectories cross only the Earth mantle,   
 characterized by an approximately constant density, $f_{\rm reg}$    
is simply given by~\cite{Lunardini:2001pb}   
 %......................................................   
 \begin{equation}   
 f_{\rm reg} = -\sin 2 {\tilde\theta}_{12} \sin(2{\tilde\theta_{12}} -2 \theta_{12})   
 \sin^2 \left(\frac{\msqstilde L}{4E}  \right) \,\ ,   
 \end{equation}   
 %..................................................   
  where ${\tilde\theta}_{12}$ is the effective value of the antineutrino    
mixing angle $\theta_{12}$ in matter,    
  $\msqstilde$ is the solar mass squared difference in matter,   
 and $L$ is the path length in Earth.   
In Earth matter, we have   
 $\sin 2\tilde\theta_{12}>0$ and $\sin (2{\tilde\theta}_{12} - 2\theta_{12})<0$,    
which tells us that $f_{\rm reg} \geq 0$.   
\begin{figure*}[t]   
\includegraphics[angle=0,width=0.7\textwidth]{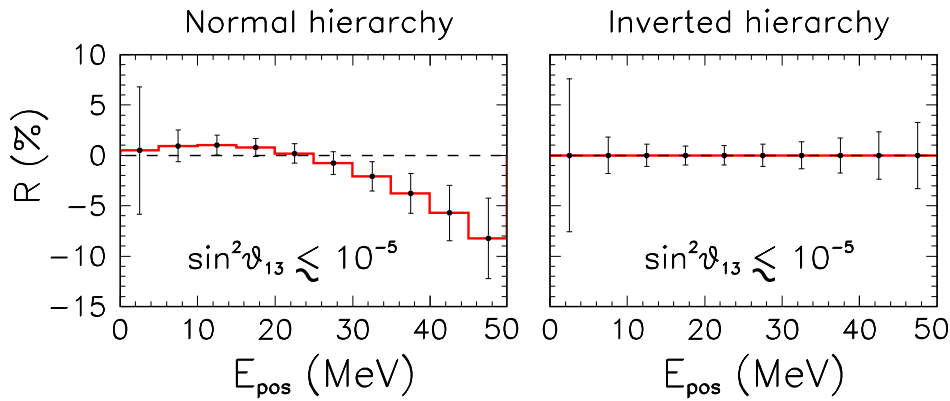}   
\caption{Plot of the ratio $R$ defined in Eq.~\ref{R-def}, as a function of    
the observable positron energy for normal hierarchy (left panel)    
and inverted hierarchy (right panel),  
with $\sin^2 \theta_{13} \lesssim 10^{-5}$.  
For $\sin^2 \theta_{13} \gtrsim 10^{-3}$, the ratio $R$ will be identical 
to the left panel for either hierarchy. 
}   
\end{figure*}

The flavor dependent primary neutrino spectra $\varphi_\nu(E)$    
can be conveniently parametrized as~\cite{Keil:2002in}   
\beq   
\varphi_\nu(E) = \frac{(\alpha+1)^{(\alpha +1)}}{\Gamma(\alpha +1)}   
\left(\frac{E}{\langle E \rangle_\nu}\right)^{\alpha}   
\frac{e^{-(\alpha+1)E/\langle E\rangle_\nu}}{\langle E \rangle_\nu} \,\ ,   
\label{def-fluxes}   
\eeq   
where $\Gamma$ is the Euler Gamma function, $\langle E \rangle_\nu$    
is the average energy for the different neutrino species,    
and $\alpha$ is the spectral pinching parameter.   
 The neutrino flux at the neutrinosphere can then be estimated to be   
%.....................................................   
\begin{equation}   
F^0_\nu = \frac{L_\nu}{\langle E \rangle_\nu} \varphi_{\nu}(E) \,\ ,    
\end{equation}   
%..................................................   
where $L_{\nu}$ is the luminosity in the $\nu$ flavor.   
All SN models robustly predict   
$\langle E_{\bar{e}} \rangle < \langle E_{\bar{x}} \rangle  
\approx \langle E_{\bar{y}} \rangle$,    
as well as $\alpha_{\bar{e}} \approx \alpha_{\bar{x}}   
\approx \alpha_{\bar{y}} $.    
This implies that the sign of  
$(F_{\bar{e}}^0 -F_{\bar{x}}^0)$    
is positive at low energies (before the crossing of   
 the $\bar{\nu}_e$ and $\bar{\nu}_x$ spectra) and negative at higher energies.   
   
The net result is that when we compare the antineutrino fluxes between  
a shadowed and an unshadowed detector,  we will have $\Delta F >0$   
 at low energies and $\Delta F<0$ at high energies in the case of normal mass hierarchy,   
or in inverted mass hierarchy with large $\theta_{13}$.    
In inverted hierarchy with small $\theta_{13}$, one expects a $\Delta F$   
compatible with zero.

To illustrate the above, we consider a galactic supernova explosion at    
a distance of 10~kpc, with luminosities   
$L_{\bar x} = L_{\bar{y}} = 0.8 L_{\bar e}$    
and total emitted energy $E_B = 3 \times 10^{53}$~erg.    
We also choose   
$\langle E\rangle_{\bar e} = 15$~MeV, $\langle E\rangle_{\bar x} =   
\langle E\rangle_{\bar y} = 18$~MeV,    
and $\alpha=3$, inspired by the results of the Garching   
 simulations~\cite{Keil:2002in,Raffelt:2003en}.   
We analyze the detection of the above signal using two large  water   
 Cherenkov detectors $A$ and $B$ of fiducial mass $0.4$ megaton each,    
as proposed for upcoming experiments~\cite{Autiero:2007zj,Jung:1999jq,Nakamura:2003hk}. We compare the number of events in detector $A$, where neutrinos arrive   
after traversing $L=8000$ km in Earth mantle with an    
approximately constant density  $\rho=4.5$~g$/$cm$^3$,   
 with another detector $B$ for which the supernova is not    
shadowed by the Earth ($L=0$).  
The reference values for features of the detectors, e.g.,    
energy resolution and interaction cross sections,  
are the same as in~\cite{Fogli:2004ff}.     
We choose $\msqs= 8 \times 10^{-5}$~eV$^2$ and $\sin^2 \theta_{12}=0.29$   
 as the oscillation parameters relevant for the Earth matter effect. 
   
We define   
\beq   
R \equiv  (N_A - N_B)/N_B \,\    
\label{R-def}   
\eeq   
as the difference between the number of $\bar{\nu}_e$  
events at the shadowed detector and the unshadowed detector,  
normalized to the number of events at the unshadowed detector.    
In Figure~1, we plot the ratio $R$ as a function of the measured positron    
energy $E_{\rm pos}$ for ${\bar\nu}_e$ in normal hierarchy (left panel)    
and inverted hierarchy (right panel)  
for $\sin^2 \theta_{13} \lesssim 10^{-5}$.  
The error bars show the statistical error in $R$.   
In the  other extreme case of $\sin^2 \theta_{13} \gtrsim 10^{-3}$,  
both the normal and inverted hierarchy would correspond to the 
left panel.

Let us consider the scenario where $\theta_{13}$ is known to be
small.
From the figure, in normal hierarchy the ratio $R$ is positive for    
$E_{\rm pos} \lesssim 25$~MeV and negative at higher energy.  
The low energy spectrum is dominated by statistical   
error, but for $E_{\rm pos} \gtrsim 30$~MeV  
the depletion of the signal with respect   
to the unshadowed detector is clearly visible, with $|R| \gtrsim 5 \%$.    
On the other hand, in inverted hierarchy we find $R=0$.   
The difference in the predictions of two hierarchies is 
significant and should be observable. 
Primary spectra taken from Livermore simulations~\cite{Totani:1997vj},   
which predict a larger difference between $\bar{\nu}_e$ and 
$\bar{\nu}_x$ average energies, would show a more  
pronounced Earth effect.   
We emphasize that our method is based on a   
model independent signature which does not rely on fitting or  
extracting any parameters.

The comparison of the neutrino signal in two detectors is also possible  
using only a single 
megaton class water Cherenkov detector together with the km$^3$  
ice Cherenkov detector IceCube at the 
South Pole~\cite{Halzen:2006mq}. 
Even though IceCube cannot reconstruct the neutrino spectrum at SN 
energies, the ratio of luminosities at these two detectors can be 
determined rather accurately, which will show about 5\% time variation  
if Earth effect is indeed present~\cite{Dighe:2003be}.  
Moreover, if a large scintillator detector~\cite{Autiero:2007zj} 
is built, 
its superior energy resolution would allow the observation of 
the modulations induced by the earth effect in the spectrum, without  
the need to compare the signal with another   
unshadowed detector~\cite{Dighe:2003jg}.

The swap of the $\bar{\nu}$ spectra due to collective effects
does not depend on the exact neutrino density profile as long
as the propagation is adiabatic \cite{Raffelt:2007cb}, 
whose validity we have checked for typical SN profiles and 
$\theta_{13}$ as low as $10^{-10}$.
Decoherence effects are highly suppressed due to the 
$\nu_e$--$\bar{\nu}_e$ flux asymmetry \cite{EstebanPretel:2007ec}, and 
other multi-angle effects also do not affect the net antineutrino
conversions substantially \cite{Fogli:2007bk}.
Moreover, with an extremely small $\theta_{13}$, the detailed matter
density profile near the $H$ resonance is immaterial, and the effects 
of density fluctuations or turbulence may safely be ignored. 
Therefore, one can make the following statements:
(i) Observation of Earth matter effects cannot be explained in 
inverted hierarchy
(ii) Nonobservation of Earth matter effects cannot be explained 
in normal hierarchy (unless the primary fluxes are almost identical).
Our proposed method is thus quite robust, and would be able to
identify the mass hiererchy.
It is not only competitive with the long baseline strategy proposed in 
\cite{deGouvea:2005mi}, but also offers an independent astrophysical 
resolution to the hierarchy determination problem.

If $\theta_{13}$ is known to be large, the hierarchy can be determined
through a number of other observables in the SN burst itself:
signatures of SN shock-wave propagation in the ${\bar\nu}_e$ signal
\cite{Fogli:2004ff,Tomas:2004gr,Fogli:2006xy},   
the $\nu_e$ signal during the neutronization burst 
\cite{Kachelriess:2004ds},
or the direct, albeit extremely challenging,
observation of the spectral split in $\nu_e$ spectrum
\cite{Duan:2007bt} at a large liquid Argon detector 
\cite{GilBotella:2003sz}.   
In fact the hierarchy may even be identified at the long baseline 
experiments.
However in such a scenario, the Earth matter effects act as an evidence 
for collective flavor conversions,
thus giving us confidence about our understanding 
of the processes happening in the core of the star.

To conclude,
determination of the leptonic mixing angle $\theta_{13}$ and  
the neutrino mass hierarchy represent two of the next frontiers  
of neutrino physics.   
In this Letter, we have proposed a new possibility for identifying   
the neutrino mass    
hierarchy that works  for extremely small values of $\theta_{13}$,    
far beyond the sensitivity of current and future terrestrial neutrino  
experiments.   
The sensitivity of supernova neutrino oscillations to the mass hierarchy, 
for    
incredibly small values of $\theta_{13}$, is a consequence   
of the collective neutrino oscillations that take place    
near the supernova core. These effects interchange the initial $\bar{\nu}_e$    
and $\bar{\nu}_y$ spectra in the inverted hierarchy, which are then    
 further processed by MSW effects in the SN envelope. 
This spectral swap can be revealed by comparing   
the event rate at a shadowed detector with that at 
an unshadowed detector.    
If neutrino oscillation experiments fail to determine  
the mass hierarchy, then this proposed method could 
represent the last hope to resolve this issue,    
provided that large  water Cherenkov detectors are available 
at the time of the next galactic SN explosion. 
This perspective should be considered when choosing optimal detector   
locations for upcoming large neutrino detectors~\cite{Mirizzi:2006xx}.

The observation of the Earth matter effects in the inverted hierarchy   
for large $\theta_{13}$ also constitutes a    
smoking gun signature of collective oscillations in a SN,   
that arise from the as yet    
unprobed neutrino-neutrino interactions.   
 It is fascinating to realize that intriguing effects like collective neutrino   
oscillations, occurring in the deep regions of an exploding supernova, produce observable  signatures at Earth and enable us to probe neutrino properties. This confirms once again the extreme importance of supernovae as laboratories for fundamental neutrino physics.     
   
%%%%%%%%%%%%%%%%%%%%%%%%%%%%%%%%%%%%%%%%%%%%%%%%%%%%%%%%%%%%%%%%%%%%%%   
   
%\begin{acknowledgments}   
We thank Georg Raffelt for valuable comments on the manuscript,   
and Andreu Esteban-Pretel, Sergio Pastor and Ricard Tom\`as for important   
criticisms on its first version. A.M. also would like to thank  
Enrique Fernandez-Martinez for interesting discussions about  
mass hierarchy determination in future laboratory experiments.  
In Munich, this work was partly supported by the Deutsche   
Forschungsgemeinschaft (grant TR-27 ``Neutrinos and Beyond''), by the   
Cluster of Excellence ``Origin and Structure of the Universe''   
(Garching and Munich) and by the European Union (contract No.\   
RII3-CT-2004-506222). In Mumbai, partial support by a Max Planck   
India Partnergroup grant is acknowledged. A.M.\ acknowledges support   
by the Alexander von Humboldt Foundation, and   
kind hospitality at the Tata Institute for Fundamental   
Research  during the development of this work.   
%\end{acknowledgments}   
   
%%%%%%%%%%%%%%%%%%%%%%%%%%%%%%%%%%%%%%%%%%%%%%%%%%%%%%%%%%%%%%%%%%%%%%   
%%%  Bibliography  %%%%%%%%%%%%%%%%%%%%%%%%%%%%%%%%%%%%%%%%%%%%%%%%%%%   
%%%%%%%%%%%%%%%%%%%%%%%%%%%%%%%%%%%%%%%%%%%%%%%%%%%%%%%%%%%%%%%%%%%%%%   

%%%%%%%%%%%%%%%%%%%%%%%%%%%%%%%%%%%%%%%%%%%%%%%%%%%%%%%%%%%%%%%%%%%%%%   
\end{document}